\documentclass[final,5p,times,twocolumn]{elsarticle}
\usepackage{amssymb}
\journal{Nuclear Instruments and Methods B}

\usepackage[ascii]{inputenc}
\usepackage[LGR,T1]{fontenc}
\usepackage[greek,english]{babel}
\usepackage{amsmath}
\usepackage{amssymb,amsfonts,textcomp}
\usepackage{array}
\usepackage{supertabular}
\usepackage{hhline}
\usepackage{graphicx}
\makeatletter
\newcommand\arraybslash{\let\\\@arraycr}
\makeatother
\newcounter{Figure}

\begin{document}

\begin{frontmatter}





%

\title{Acceleration with Self-Injection for an All-Optical Radiation Source at LNF}

\author{L. A. Gizzi $^{a,b)}$} 
\ead{la.gizzi@ino.it}
\author{M. P. Anania $^{c)}$}
\author{G. Gatti $^{c)}$}
\author{ D. Giulietti$^{a,b,d)}$}
\author{G. Grittani$^{a,b,d)}$}
\author{ M. Kando$^{e)}$}
\author{M. Krus$^{f)}$}
\author{ L. Labate$^{a,b)}$}
\author{ T.~Levato$^{a,f,g)}$}
\author{Y. Oishi$^{(g,b)}$}
\author{F. Rossi$^{i)}$}

\address{$^{a)}$ILIL, INO-CNR, Via Moruzzi,1 56124 Pisa, Italy
\\ $^{b)}$INFN, Sez. Pisa, Largo B. Pontecorvo, 3 - 56127 Pisa, Italy
\\ $^{c)}$Laboratori Nazionali di Frascati, INFN, Via E. Fermi,  Frascati, Italy 
\\ $^{d)}$Dipartimento di Fisica E. Fermi, Universit\'a di Pisa, Italy 
\\ $^{e)}$Japan Atomic Energy Agency (JAEA) Kyoto, Japan
\\$^{f)}$Fyzik\'aln\'i \'ustav AV \v{C}R v.v.i., Praha, Czech Republic
\\$^{g)}$U. Tor Vergata, Roma, Italy
\\ $^{h)}$CRIEPI, Kanagawa, Japan
\\ $^{i)}$Universit\'a di Bologna e Sez. INFN, Bologna, Italy}
%


\begin{abstract}
We discuss a new compact ${\gamma}$-ray source aiming at high spectral density, up to two orders of magnitude higher than currently available bremsstrahlung sources, and conceptually similar to  Compton Sources based on conventional linear accelerators.  This new source exploits electron bunches from laser-driven electron acceleration in the so-called self-injection scheme and uses a counter-propagating laser pulse to obtain X and ${\gamma}${}-ray emission via Thomson/Compton scattering. The proposed experimental configuration inherently provides a unique test-bed for studies of fundamental open issues of electrodynamics. In view of this, a preliminary discussion of recent results on self-injection with the FLAME laser is also given.  

\end{abstract}


\end{frontmatter}



\section{Introduction}\label{sec::intro}
The impressive progress of high power laser technology initiated by the introduction of the Chirped Pulse Amplification (CPA) concept \cite{Strickland_OC85} is now leading to the realization of new large laser systems within the framework of the Extreme Light Infrastructure (ELI) that, by the end of this decade, will start paving the way to the exploration of new physical domains, approaching the regime of electron-positron pair creation and the possibility to reach the critical field of quantum electrodynamics\cite{Schwinger:1951aa}.  At the same time, the control of ultra-high gradient plasma acceleration \cite{malka:nature, Leemans_GeV, Giulietti_PP02} is being pursued and advanced schemes are being proposed for the future TeV linear collider \cite{leemans:3}. 

Meanwhile, existing laser-plasma accelerating scheme are being considered for the development of novel radiation sources.  All-optical X-ray free electron lasers (X-FEL) are already being explored \cite{gallacher:093102} with encouraging chances of success in the short term. All-optical, laser-based bremsstrahlung X-ray and ${\gamma}${}-ray sources  have already been explored \cite{Gizzi_PRL96, Gizzi_LPB01} and successfully  tested using self-injection electron bunches \cite{Giulietti_PRE01, giulietti_prl08} showing high efficiency and potential for laboratory applications. 

However, in order to positively enter the domain of nuclear applications, significantly higher energy and spectral density ${\gamma}$-rays are required. In this scenario, a very demanding application is the Nuclear Resonance Fluorescence (NRF) due to the small spectral width of nuclear resonance transitions. New sources have been developed \cite{PhysRevSTAB.13.070704} based upon the use of high energy LINACS and high power lasers or free electron lasers\cite{Sandorfi_Compton_83} to generate ${\gamma}$-rays via Thomson/Compton scattering.  In this scenario, the use of laser-plasma accelerated electrons has also been explored \cite{NRF_LPA_IEEE} and is regarded as a possible way to make nuclear sources far more accessible than current Linac based sources.

In a pioneering experiment \cite{PhysRevLett.96.014802}  carried out at the Jena laser Facility in 2006, all-optical Thomson scattering (TS) in the 1keV X-ray region was demonstrated using a compact configuration with a relatively low degrees of freedom for optimization and using poor quality laser-accelerated electron bunches, still affected by 100\% energy spread. Since then, laser-plasma acceleration has seen dramatic advances and laser accelerated electrons can now exceed 1GeV with energy spread well below 10\%, with record values close to 1\%. Moreover, new schemes are being proposed to control injection and optimize acceleration, which are now being implemented to further improve the quality of laser accelerated electrons.
\section{Thomson scattering source \label{sec::Thomson}}
An all-optical scheme is proposed here in which a laser-driven electron accelerator based on the design of the self injection test experiment (SITE) \cite{Gizzi_NC} is used to deliver electron bunches required to generate ${\gamma}$-rays in a Thomson back-scattering configuration\cite{Tomassini:2005aa}.
TS from free electrons is a pure electrodynamics process in which each particle radiates while interacting with an electromagnetic wave. From the quantum-mechanical point of view TS is a limiting case of the process of emission of a photon by an electron absorbing one or more photons from an external field, in which  the energy of the scattered radiation is negligible with respect to the electron's energy. If the particle absorbs only one photon by the field (the linear or non relativistic quivering regime), TS is the limit of Compton scattering in which the wavelength $\lambda_X$ of the scattered photon observed in the particle's rest frame is much larger than the Compton wavelength $\lambda = h/m_ec$ of the electron.  Since $\lambda_c/\lambda_{X}<<1$, the TS process can be fully described within classical electrodynamics both in the linear and nonlinear regimes. 

TS of a laser pulse by energetic counter-propagating electrons was initially proposed in 1963 \cite{PhysRevLett.10.75, PhysRev.138.B1546} as a quasi monochromatic and polarized photons source. With the development of ultra intense lasers the interest on this process has grown and the process is now being exploited as a bright source of energetic photons from UV to ${\gamma}$-rays and atto-second sources in the full nonlinear regime. 

\begin{figure}[h]
\begin{center}
\mbox{\includegraphics[width=0.70\columnwidth]{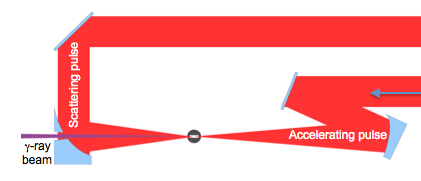}}
\caption{Schematic view of the counter-propagating configuration for all-optical X/${\gamma}$-ray generation. The main FLAME laser pulse is split in two pulses which are focused in the proximity of a gas target. One of the two pulses (right) accelerates electrons. The other pulse is focused on the accelerated electron bunch to scatter off radiation which is emitted along the electron bunch propagation direction.}
\label{fig::counterpropagating}
\end{center}
\end{figure}

As shown in Figure \ref{fig::counterpropagating}, the proposed source is based upon a counter-propagating configuration of two ultraintense laser pulses focused in a gaseous target. The counter-propagating configuration is obtained by splitting the main laser pulse in two pulses with controlled energy and independent focusing configuration.  In this configuration, one of the pulses propagates towards the electron bunches generated by the other pulse.

The other, counter-propagating pulse interacts with the energetic electrons generating radiation along the bunch propagation direction.  An additional laser pulse, the auxiliary "probe" pulse of the FLAME system is then transported to the target chamber and is used to diagnose the plasma density before and during the interaction. 

\subsection{Source parameters  \label{sec::source_parameters}}
\begin{figure}[h]
\begin{center}
\mbox{\includegraphics[width=0.70\columnwidth]{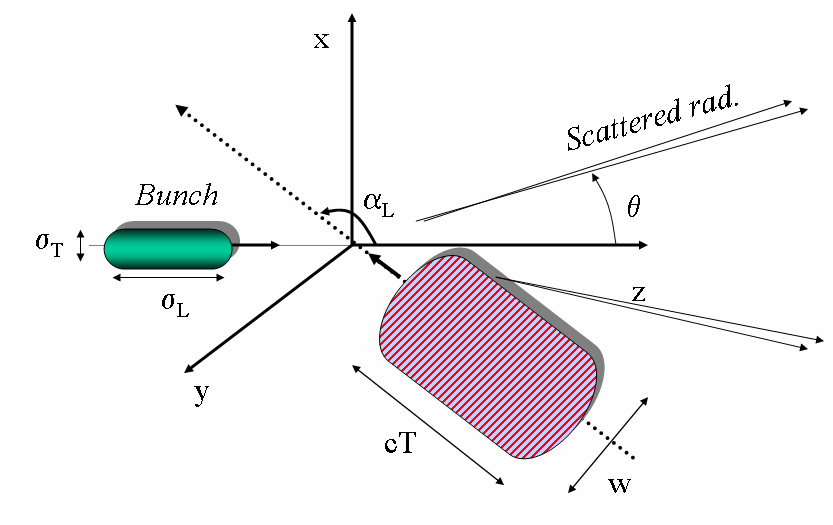}}
\caption{Thomson scattering geometry. The scattered radiation is emitted along the z axis, in a small cone of aperture $1/{\gamma}_0$. When ${\alpha}_L=\pi$ the backscattering geometry occurs. }
\label{fig::TS_geometry}
\end{center}
\end{figure}

The three main parameters of the Thomson scattering of a laser pulse by a free electron are the particle energy ${\gamma}_0$, the angle ${\alpha}_L$ between the propagation directions of the pulse and the electron and the laser pulse normalized amplitude $a_0=8.5 \times 10^{-10} \sqrt{I\lambda^2}$ where $\lambda$ is the laser wavelength in $\mu$m and $I$ is the laser intensity in W/cm$^2$. The pulse amplitude $a_0$ controls the momentum transferred from the laser pulse to the electron, i.e. the number of photons of the pulse absorbed by the electron. If $a_0<<1$ only one photon is absorbed and the quivering is non-relativistic (linear Thomson scattering). For an electron initially moving with ${\gamma}_0 >>1$ the resulting scattered radiation is emitted forward with respect to the electron initial motion within a cone of aperture $1/{\gamma}_0$ and is spectrally shifted at a peak wavelength given by:
\begin{equation}
\lambda_{X,\gamma} \simeq \lambda \frac{1-\beta \cos \theta}{1 - \beta \cos \alpha_L}
\end{equation}
where $\beta=v/c$ is the particle velocity and $\theta$ and $\alpha_L$ are defined according to the geometry of Figure \ref{fig::TS_geometry}.
Among the possible interaction geometries, the case of backscattering ${\alpha}_L=\pi$ is the most suitable for a source as it produces radiation with the highest energy:
\begin{equation}
E_{Back}\simeq 4 \gamma^2 E_0,
\label{TS_Energy}
\end{equation}
where $E_0$ is the energy of laser photons. According to Equation \ref{TS_Energy}, for a laser wavelength of 0.8 $\mu$m, the electron energy required to achieve photons with energy of 50 keV, 500 keV and 1 MeV are 46 MeV, 145 MeV and 205 MeV respectively, which are  within the accessible electron energy of laser-plasma acceleration with self-injection.  Moreover, the head-on configuration allows the highest overlap of the electron beam and the pulse and minimizes spurious effects induced by the transverse ponderomotive forces of the laser pulse. 

In the nonlinear regime ($a_0{\ge}1$) the  strong exchange between the laser pulse and particle momentum induces a relativistic electron motion, consisting of a drift and a quivering having both longitudinal and transverse components with respect to the pulse propagation. In turn, the time dependent longitudinal drifting results in a non harmonic electron motion, thus producing scattered radiation with a complex spectral distribution too. If the electron interacts with a laser pulse with a flat-top temporal shape, the spectral distribution of the scattered radiation is organized in equally spaced harmonics. The peak energy of each Nth harmonics depends now on both the energy of the particle and the pulse normalized amplitude $a_0$ and is given by $E_N \sim N 4 \gamma_0^2 E_0 (1+1/2a_0^2)^{-1}$.

It can be shown that the minimum bandwidth of a $\gamma$-ray beam produced by a back-scattering source depends upon the electron energy spread $\Delta \gamma_0/\gamma_0$, the electron rms transverse emittance $\epsilon_n$,  the beam spot size at the collision point $\sigma_x$, the reduced laser intensity $a_0$ and the laser spectral bandwidth $\Delta \nu_L/\nu_L$ according to the equation: 
\begin{equation}
\Delta \nu_{\gamma}/\nu_{\gamma} \simeq 2 \Delta \gamma_0/\gamma + 2(\epsilon_n/\sigma_x)^2+ a_0^2(1+a_0^2) + \Delta \nu_L/\nu_l
\label{eq::spread}
\end{equation}
Taking typical values of a self-injected electron beam, as those currently produced in the bubble regime \cite{bubble_apb, gord_pop}, at the exit of the plasma, we have that the first two terms in Equation \ref{eq::spread} are in the range 1-10\%, the laser bandwidth is typically of the order of 10\%. A collision laser pulse with  $a_0<1$ will then be required to avoid further increase of the  bandwidth for the ${\gamma}${}-ray beam. Overall, we can therefore expect a bandwidth in the range of 5-10\% at the best for an optimal arrangement, corresponding to 500 keV to 1 MeV bandwidth for 10 MeV photons. 
In these circumstances, the rate of photon generation is given by 
\begin{equation}
N_{\gamma}=2.1 \times 10^8 U_L[J]\  Q[pC] \ h\nu^{-1}[eV]\  \sigma^{-2}_x[\mu m] \ f,
\end{equation}
where $U_L[$J] and $Q$[pC] are the laser pulse energy and the electron bunch charge.  Assuming a typical value for the bunch charge  $Q=1000$ pC, a repetition rate $f=10$ Hz, a matched focal spot diameter $\sigma_x=8 \ \mu$m and a laser pulse energy $U_L=2 $ J, we obtain $N_{\gamma}=3.5 \times 10^{10}$ photons/sec over the entire solid angle and spectral bandwidth. This corresponds to approximately $10^8$ photons/sec within a 10\% bandwidth. Finally, based on current optimum performance of acceleration with self-injection, spectral intensity at 1 MeV bandwidth would be approximately $10^2$ photons/sec/eV. 

If we consider other laser-plasma acceleration schemes, e.g. those based upon capillary discharge gas targets, we can expect a narrower bandwidth and lower transverse emittance \cite{Wiggings_PPCF_2010}. This is done at the expenses of the charge available in the accelerated bunch, typically in the few pC range. This is a crucial aspect of the proposed experiment and will require further development.

\section{Towards high fields effects.}
In the general description of the interaction of a charged particle with an external e.m. field, the emission of radiation by the accelerated particle gives rise to a back-reaction, the so called Radiation Reaction (RR) also referred to as radiation friction.  In the usual approach, the problem is solved in steps. In the first step, the motion of an electron is calculated using the Lorentz force for the given external EM fields. In the second step, the radiation emitted by the electron can be calculated given its motion. This two-step process is non self-consistent because it neglects the back-reaction on the electron by the EM fields generated by its motion. To make the electron motion consistent with the emission of radiation which carries away energy, momentum and angular momentum, it is found that an additional term, the RR force, must be added to the Lorentz force. 

However, in most cases, the non self-consistent approach still provides a very accurate description of the particle dynamics. This is the case for example, of particle accelerators and of most laboratory and astrophysical plasmas. These conditions are preserved provided that the energy radiated by the particle  is small compared with the energy of the system. 
It can be shown that in the case of a charged particle oscillating in an external field, radiation dominates when the motion of the particle changes appreciably in a time $t=10^{-24}$ s, i.e. over a distance $ct=10^{-13}$ cm.  Interestingly, the latter distance is comparable with the classical electron radius. These considerations immediately tell us that the possibility of accessing this regime in the laboratory is extremely challenging. Ultraintense, femtosecond laser pulses are regarded as a possible tool to enter this regime and future laser systems currently under construction are expected to make this opportunity even more realistic. 

The question of a description of the RR force is a long-standing, highy controversial and, to a large extent, still unresolved, facing issues of electrodynamics that ultimately deal with the actual structure of the electron charge, its nature and the role of quantum effects \cite{jd_jackson}. The perspective of probing RR effects with ultra-intense lasers has further stimulated the interest in these problems and  alternative models to describe RR have been proposed, revitalizing the long standing debate. Probably the simplest, but already effective approach is that of Landau and Lifshitz (LL) \cite{Landau_Lifshitz} which is free from pathologies apparent in other models. Currently, the effort is not limited to a rigorous formulation of the RR force on the single electrons but it is also devoted to understand how RR may be implemented consistently and effectively in a many-body system, i.e. in a plasma. Finally, experimental validation is needed to discriminate among available theoretical and numerical models and a road-map for this validation is now being established at the main high-power laser facilites world-wide.
\subsection{Seeking experimental evidence of Radiation Reaction\label{sec::exp_evidence}}
A theoretical study \cite{koga:093106} of the motion of an ultrarelativistic electron in an ultraintense EM field, based on the LL equation, has characterized a Radiation dominated regime as the system in which the energy gain by the electron equals the radiation loss. As anticipated above, it can be shown that this condition occurs when 
the dimensionless field amplitude $a_0 >  400$
which implies a laser intensity of 
$I_L > 10^{24}$ W/cm$^2.$
This is a very high laser intensity, 3 orders of magnitude above current capability of existing laser systems. However, ultraintense laser technology is providing increasingly high electromagnetic field intensities. It is therefore foreseeable that the next generation of high power laser systems will allow this regime to be accessed relatively soon.
On the other hand, according to recent models based upon LL equations, it is predicted that experimental evidence of RR effects can be obtained in Thomson scattering configurations at relatively lower laser intensities, below the foreseen threshold for the RR-dominated regime, and not far from the maximum intensities available from PW-class laser systems. In addition, other models \cite{Bulanov_Flying_Mirror} predict laser pulse intensification and shortening in a self-injection laser wakefield acceleration configuration which could enhance the effect and make RR accessible at existing laser facilities. 

In a recent theoretical paper \cite{PhysRevLett.102.254802}, the interaction between a superintense laser pulse at $5 \times 10^{22}$ W/cm$^2$ and a 40 MeV counter-propagating electron was investigated looking at the effect of Radiation Reaction on the spectrum of Thomson Scattered radiation. 
According to this study, the angle and frequency resolved spectra show signatures of RR dependent effects on the angular distribution of scattered radiation. 
These studies confirm that, in principle, anomalies in the Thomson emission due to RR are not far from the capabilities of current laser systems. In view of this, feasibility studies towards the experimental realization may be already investigated, starting from a dedicated, start-to-end simulation of the entire interaction configuration. Moreover, from an experimental viewpoint, control of the laser-plasma acceleration process is necessary to establish parameters of the accelerated electron bunch to be included in the simulations.

From the modelling viewpoint, this approach uses numerical tools capable of describing the dynamics of electron beam acceleration and interaction with the counter-propagating pulse in a realistic geometry and incorporating RR effects. Recently, several self-consistent simulation studies incorporating RR in laser-plasma interactions via the LL force in a Particle-In-Cell (PIC) code have been performed (see e.g. \cite{Tamburini_NJP_2012}). Numerical efficiency is of a paramount importance for such simulations (which require large supercomputers) because in a PIC code the run time mostly depends of the calculation of particle acceleration, hence on details of the force term. The numerical implementation of Ref.\cite{Tamburini_NJP_2012} proved to be efficient enough to allow fully three-dimensional simulations and was thus suitable for full-scale simulations for the experiment proposed in \cite{PhysRevE.85.016407}. A similar approach is being followed for the full modelling of the proposed radiation source.

\section{Acceleration with Self-injection at FLAME\label{sec::self_injection}}
In the framework discussed above, the  FLAME laser system has recently been commissioned at LNF and experimental runs dedicated to laser-plasma acceleration experiment with self-injection (SITE) have already been carried based on previous pilot experiments \cite{Channeling_2008_book}. The first run was carried out during the early commissioning stage in 2010, at relatively low laser power ($\approx$10 TW), and showed successful generation of mono-energetic, high energy electron bunches with a moderate to high degree of collimation down to the 6 mrad level \cite{Tadzio_Varenna_School_2011}. A second experiment at higher laser power ($\approx$100 TW) was carried out in July 2012 and enabled us to further explore the planned experimental configurations \cite{Gizzi_NC, TDR_2009}. A detailed description of these results will be given elsewhere.  Here give an overview of the experimental set up with some  preliminary highlights of the results, with focus on the stability and the reproducibility of the observed acceleration process in view of the application to the proposed TS radiation source.
\begin{figure}[h]
\centering
\includegraphics[width=0.45\columnwidth]{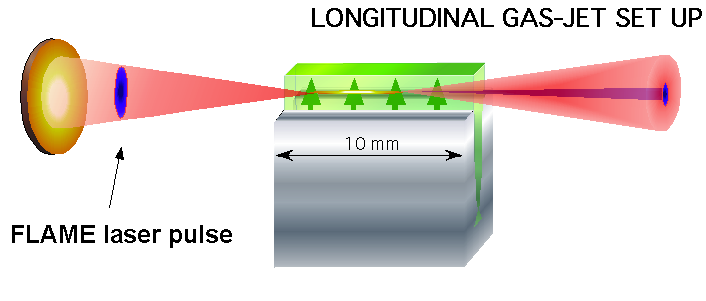}
\includegraphics[width=0.45\columnwidth]{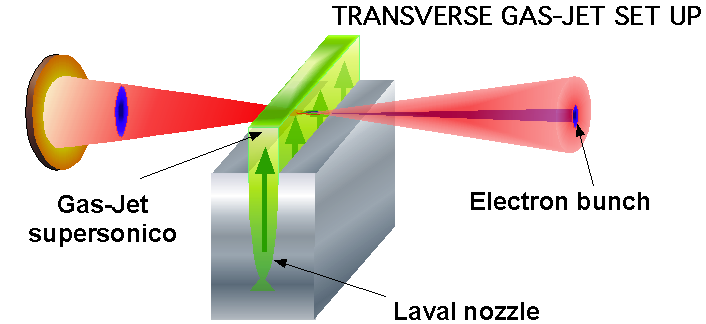}
\caption{Schematic layout of the laser-gas-jet interaction for the production of energetic electron bunches in the self-injection configuration used in the SITE experiment at LNF. We used two different gas-jet length of 10 mm (longitudinal propagation) and 1.2 mm or 4 mm (transverse propagation).}
\label{fig::nozzle_configs}
\end{figure}

In the self-injection scheme proposed here, electron bunches are generated from laser-plasma interaction with a rectangular gas-jet of a few millimeters in the so-called bubble regime \cite{bubble_apb, gord_pop}. In this regime, a short ($c \tau < \lambda_p/2$) and intense ($a_0$ {\textgreater} 2) laser pulse rapidly ionizes the gas \cite{Gizzi_PRE_06, Gizzi_PRE_09} and expels the plasma electrons outward creating a bare ion bubble. The blown-out electrons form a narrow sheath outside the ion bubble and the space charge generated by the charge separation pushes the electrons back creating a bubble-like wake. For sufficiently high laser intensities ($a_0 \ge$ 3-4) electrons at the back of the bubble can be injected in the cavity, where the longitudinal accelerating field is of the order of \ $100 \ \Delta n $ [cm$^{-3}] $V/m, where $\Delta n$ is the amplitude of the local electron density depression in the wake.

The FLAME laser meets both conditions of short pulse duration and high intensity required to achieve this condition. When the laser pulse impinges onto the gas-jet it promptly excites (without significant pulse evolution) a bubble wake where electrons are readily injected leaving the almost entire gas-jet length for the acceleration process. A proper choice of plasma and laser parameters to ensure an optimized acceleration process can be obtained using  a phenomenological theory \cite{wei-lu} to account for dephasing and depletion of the laser pulse.  The basic working point currently under consideration for the self-injection configuration at FLAME is the one described in Table 1. In this case, following the phenomenological description, at the optimum laser performances we can expect a quasi monochromatic (few \% momentum spread) bunch with a charge of ${\approx}$ 0.6 nC and an energy of approximately 1.0 GeV after 4 mm propagation.
\begin{figure}[h]
\centering
\includegraphics[width=1\columnwidth]{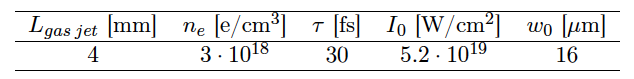}
\caption{A possible working point of the self-injection test experiment at FLAME for laser-acceleration in the GeV scale.}
\label{fig::nozzle_configs}
\end{figure}

This scenario is confirmed by 3D PIC simulations performed with the fully self-consistent, relativistic, electromagnetic PIC code ALaDyn \cite{aladyn_ieee, carlo-NIMA}. Infact, simulations predicts a bunch with an energy of 0.9 GeV, a momentum spread (rms) of 3.3\%, a bunch charge of 0.6 nC, a bunch length of 1.8 {\textmu}m (the average current is 50 kA) and a beam divergence (rms) of 2.8 mrad. 

\begin{figure}[h]
\centering
\includegraphics[width=0.46\columnwidth]{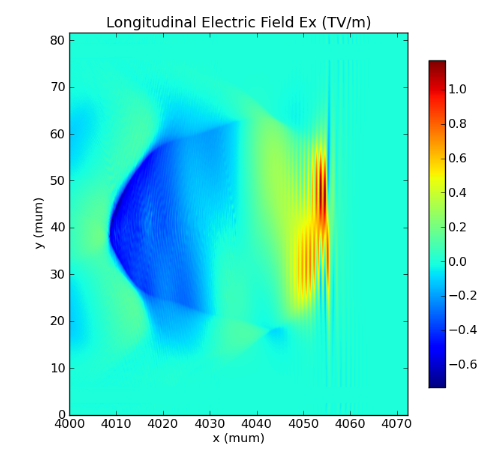}\ \ \ 
\includegraphics[width=0.45\columnwidth]{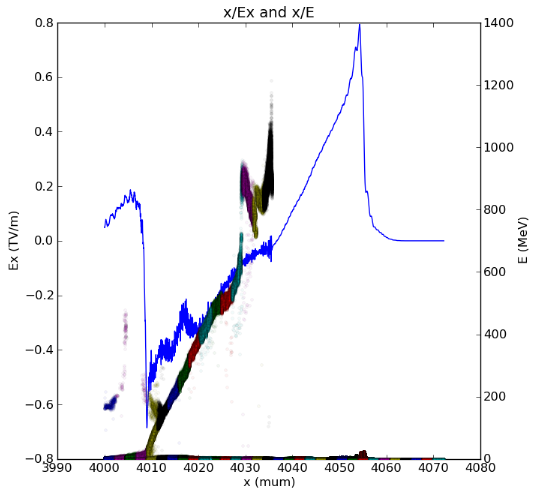}
\caption{Simulation results of the numerical code JASMINE showing the output for self-injection for optimized performance of the FLAME laser system and a gas-jet intensity of 3$\times 10^{18}$ cm$^{-3}$. on a 4 mm gas-jet.} 
\label{fig::jasmine}
\end{figure}

More recently these results were basically confirmed by simulations carried out using the 3D GPU particle in cell code. 
In fact, advanced numerical tools in the modelling scenario use computer architectures based upon graphical processing units (GPU) and are proving to provide much faster simulations \cite{Rossi_AAC_2012}. Indeed, by using a 32-GPU cluster it was possible to perform numerical simulations of a 4 mm gas-jet in approximately two days, reaching the same accuracy as previously obtained with the code Aladyn \cite{aladyn_ieee}. The plots of Figure \ref{fig::jasmine} show the longitudinal field (left) and the electron energy gain (right) for the full SITE case of laser intensity of 5$\times 10^{19}$ W/cm$^{2}$ on a matched plasma electron density of 3$\times 10^{18}$ cm$^{-3}$. 
As shown in Figure \ref{fig::jasmine}(left) the accelerating field exceeds the 0.3 TV/m in an accelerating structure of approximately 30 $\mu$m in diameter. The electron energy after the 4 mm acceleration peaks at approximately 830 MeV, with an energy spread of 6\%.  In addition, a low energy component is also visible as a well separated component. The simulation also yields the angular divergence and the charge of the high energy component which are found to be approximately 8 mrad and 0.6 nC respectively.

From an experimental perspective, the working point given above requires the optimized performance of the FLAME laser system, with special attention to the transverse phase and, therefore, to the quality of the focal spot. This is of a particular concern when operating the laser at the maximum output laser energy of 7 J and will require installation of an adaptive optics. In the mean time, preliminary experimental runs were carried out a maximum output laser laser energy of 4 J. Up to this energy level of energy no significant phase front distortion was found to occur and the Strehl ratio was measured to be > 65 \%. Taking the flat top beam size of 90 mm diameter (corresponding to a 120 mm diameter aperture) and assuming an  $M^{ 2}=1.5$,  we have a maximum nominal laser intensity on target of 2$\times 10^{19}$ W/cm$^2$.

\section{First experimental run on self-injection  \label{sec::results_site}}
\begin{figure}[h]
\centering
\includegraphics[width=0.3\columnwidth]{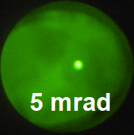}\ \ \ \
\includegraphics[width=0.32\columnwidth]{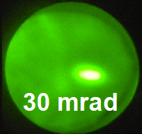}
\caption{Images of the accelerated electron bunch on a scintillating screen (Lanex) showing the production of collimated electron bunches of minimum (5 mrad) and maximum (30 mrad) divergence. Images were recorded during the 1st SITE run.}
\label{fig::site_first_lanex}
\end{figure}

\begin{figure}[h]
\centering
\includegraphics[width=0.5\columnwidth, height=3cm]{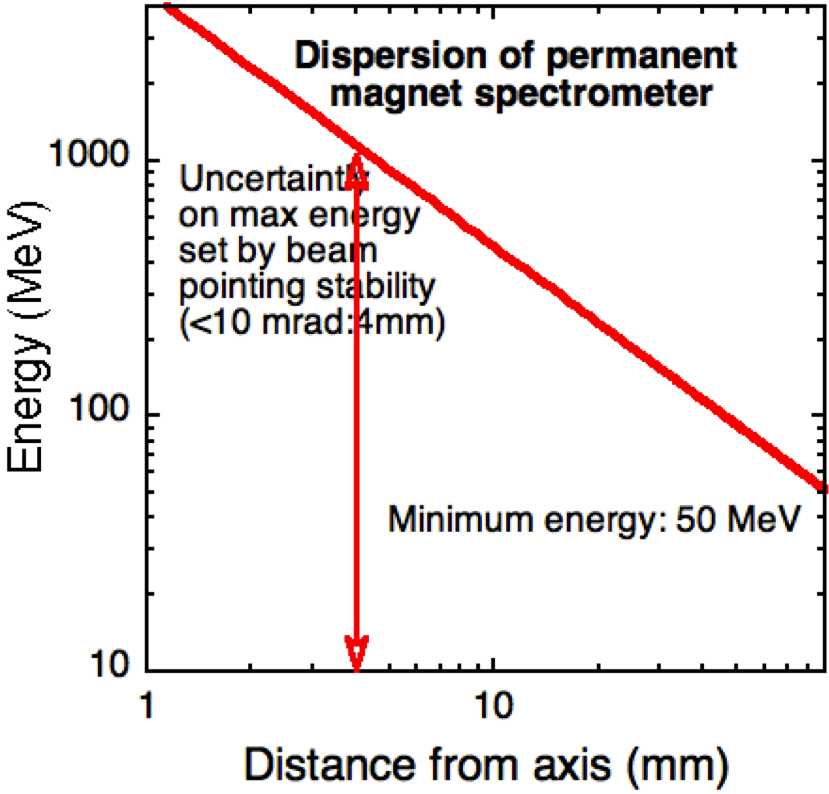}
\caption{Dispersion curve of the permanent magnet spectrometer showing the energy of electrons vs. landing position on the Lanex scintillating screen. The vertical arrow indicates the minimum detectable deflection.}
\label{fig::magnet_dispersion}
\end{figure}

A stable regime of production of collimated bunches was established during the first SITE run in 2010 using the transverse gas-jet configuration of Figure \ref{fig::nozzle_configs} with nitrogen, working with a back pressure of about 17 bar corresponding to a maximum gas density of approximately $1\times 10^{19}$ atoms/cm$^3$.  These preliminary data were obtained using an F/10 focusing optics, at a fixed pulse duration of about 30 fs and for an energy per pulse ranging from a minimum of about 300 mJ up to 1 J, corresponding to a laser intensity on target ranging from  $ 3\times 10^{18}$\ W/cm$^2$ to a maximum of $ 9 \times 10^{18}$\ W/cm$^2$.  Figure \ref{fig::site_first_lanex} shows a sample of the data in which electron bunches with a divergence in the range between 5 and 30 mrad were systematically accelerated. 
%
%
%
%
%
Preliminary electron energy measurements obtained initially with a stack of radiochromic films using a technique discussed in \cite{sheeba} confirmed that the energy of electron accelerated at 1 J  laser energy were already in the 100 MeV range. In addition, a spectrometer consisting of a magnetic dipole was used to obtain a more quantitative estimate of the electron energy. It was based upon a couple of permanent magnets mounted on a C-shaped iron structure to form a closed magnetic loop with a 5 mm gap with a quasi uniform magnetic field of 1 T. Electrons were set to propagate across this magnetic field region where are deflected according to their energy and then land on a  position on the Lanex screen according to the plot of Figure \ref{fig::magnet_dispersion}. According to this plot, electrons of  50 MeV experience a deflection of approximately 80 mm on the screen and this deflection becomes 10 mm at 500 MeV.  

By using this spectrometer it was possible to observe the main spectral features of the accelerated electrons, increasing the laser energy at fixed pulse duration and focal spot.  Analysis of these spectra showed that the highly collimated electron bunches have a typical electron energy of 100 MeV,  consistent with the value measured using the RCF technique. 
Data were also taken at higher laser energy, up to the value of 2.5 J before compression. These additional measurements enabled us to identify the role of phase front distortions affecting the laser pulse during the first FLAME commissioning phase. In fact, further increasing the laser energy above the 1 J level showed evidence of beam break up in the focal region which prevented laser intensity to increase as expected. As discussed below, these issues were identified and partially corrected during the second SITE run. 

\section{Second experimental run on self-injection  \label{sec::results_site}}
\begin{figure}[h]
\centering
\includegraphics[width=0.7\columnwidth]{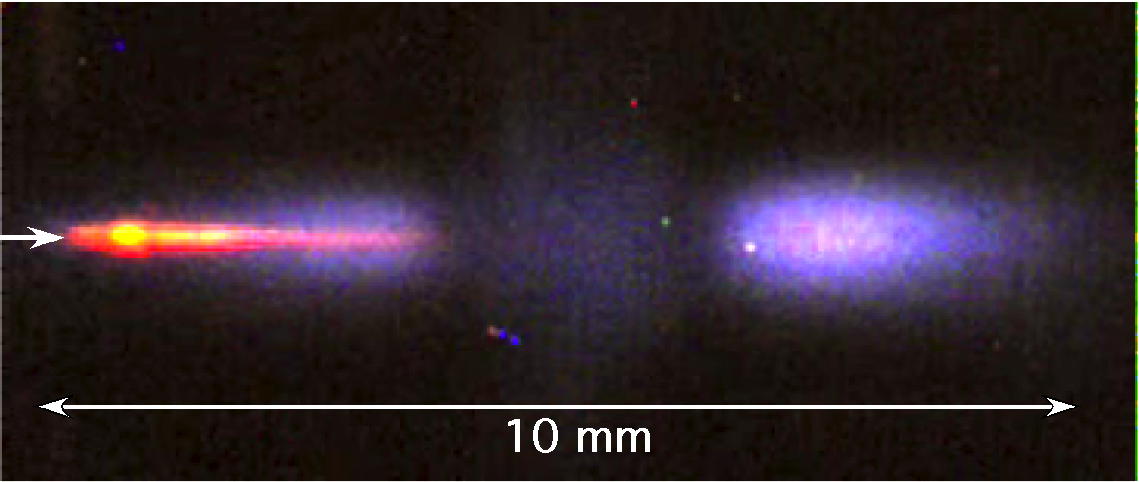}
\caption{Images of the Thomson emission from propagation of the laser pulse in a Nitrogen gas-jet at 10 bar valve pressure. The laser propagates from left to right. Also visible in the image is the plasma self-emission.}
\label{fig::thomson}
\end{figure}

A second SITE run took place in July 2012 and was focused on a more systematic study of the acceleration process for different gas type, gas density using both the transverse and longitudinal gas-jet configurations of Figure \ref{fig::nozzle_configs}. We used nitrogen and helium with a back pressure ranging from 1 to 17 bar, corresponding to a maximum gas density approximately in the range from $6\times 10^{17}$ and $10^{19}$ atoms/cm$^3$ and a maximum  laser-intensity on target of $2\times 10^{19} $\ W/cm$^2$.  A total of 3000 shots with e-bunch production was recorded and data were taken with an optical imaging system and Lanex screen placed at 475 mm from the gas-jet to measure both the electron bunch transverse size and, with the insertion of the 50 mm long magnetic dipole at 132.5 mm from the gas-jet, the energy spectrum. In addition, a shadowgraph of the image was taken to measure the longitudinal extent and the transverse size of the interaction region. The imaging system was set to view the interaction region in the vertical direction, i.e. along the axis perpendicular to the laser polarization plane to image out the Thomson scattered radiation. The plot of Figure \ref{fig::thomson} shows the typical Thomson image obtained from measurements with the longitudinal gas-jet configuration. As shown by the arrow, the laser propagates from left to right showing a clear Thomson emission (red-yellow in the color figure)  in the first 3.2 mm (FWHM) of the propagation. Taking into account the laser beam parameters, we expect a depth of focus of approximately $\pm $ 260 $\mu$m. Therefore, according to the image of Figure \ref{fig::thomson},  we find that laser propagation occurs over a propagation length which is several times the depth of focus. 

Propagation length was found to be dependent on the gas density and pressure, ranging from approximately 1.3 mm (FWHM) for 70\% Nitrogen gas mixture (air) at 1 bar pressure to less than 2 mm (FWHM) for Helium at 15 bar pressure. These results indicate that propagation length increases at higher electron density where stronger refraction effects may occur on the  propagation of the laser pulse in the plasma. At this stage we can anticipate possible contribution of self-focusing effects to the observed behavior of the laser pulse. In fact,  according to the well known expression for the critical power for relativistic self-focusing, $P_{cr} \approx 17 (\omega / \omega_{p})^2$ GW and taking into account the estimated maximum electron density given above, we find that the critical power in our experimental conditions ranges from $ 3  $ TW for the highest density case to approximately 50 TW of the lowest density case. In the case of 1J of laser energy on target, corresponding to a laser power exceeding 30TW, we expect  the interaction in the higher density case to be affected by self-focusing that could set the conditions for a moderate channeling of the laser pulse, thus effectively extending the propagation length. Detailed numerical simulation will be necessary for a confirmation of this result.  

\begin{figure}[h]
\centering
\includegraphics[width=0.4\columnwidth]{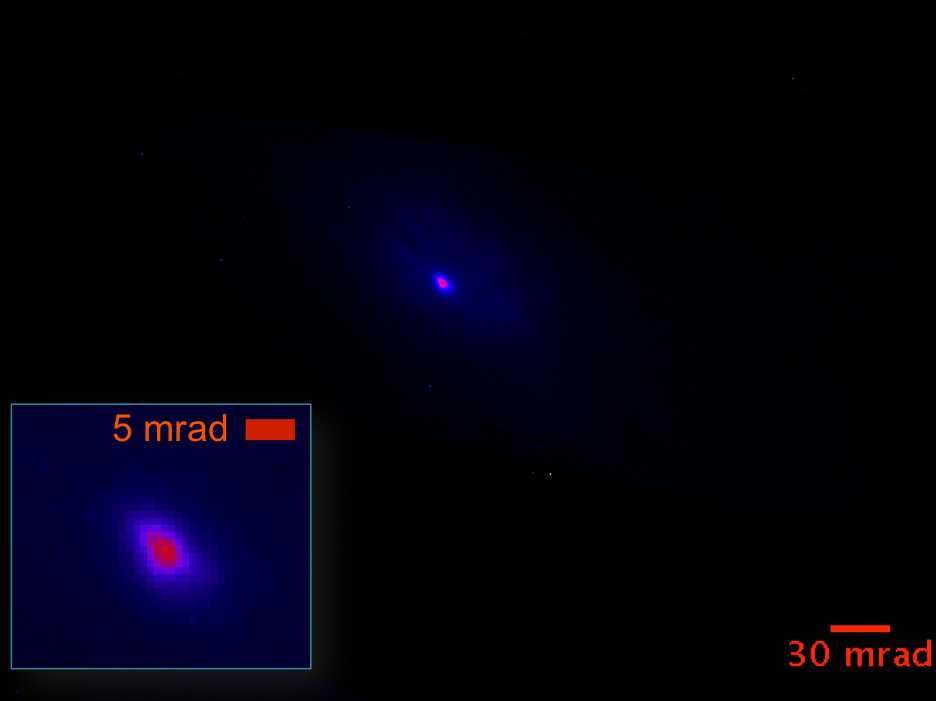} \ \ \
\includegraphics[width=0.4\columnwidth]{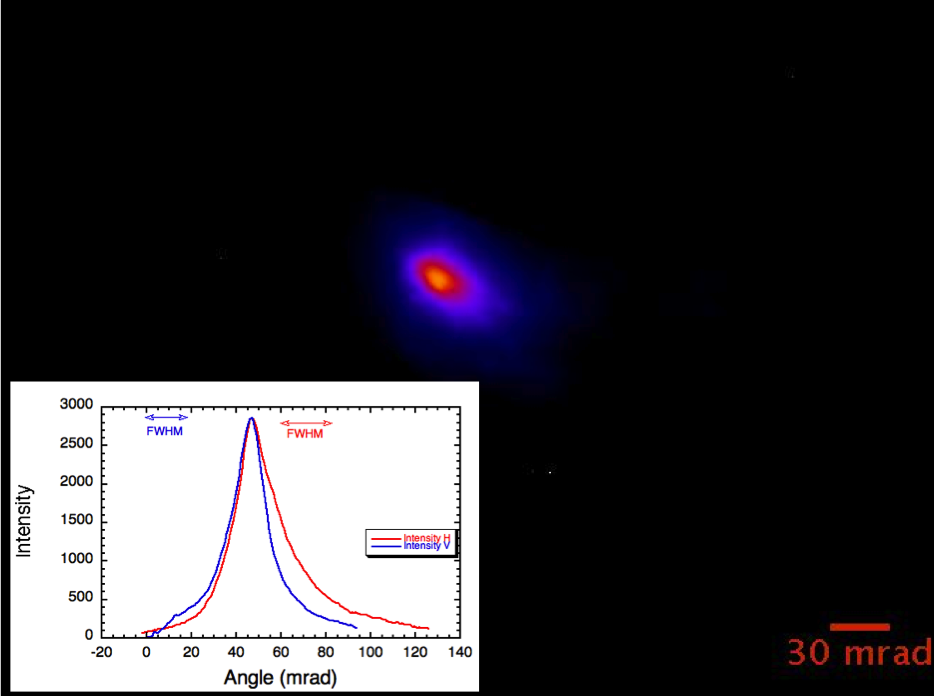}
\caption{Left: Image the scintillating LANEX screen at 47 cm downstream the interaction point showing the bunch transverse size of approximately 0.5 mm and a corresponding bunch divergence of 1 mrad.  Right: integrated image of the electron bunch over 30 laser shots showing a total pointing stability of approximately 10 mrad. The bottom-left plot shows the lineout of the image across the vertical and horizontal directions.}
\label{fig::divergence}
\end{figure}

Information about the accelerated electron bunches was obtained by using the LANEX scintillating screen to measure both the angular divergence and the energy spectrum. The image of Figure \ref{fig::divergence} (left) shows the typical image of a single electron bunch obtained from optimized acceleration in Nitrogen. According to this image, the single bunch exhibits a divergence of approximately 1 mrad FWHM, a value significantly smaller than that measured during the first SITE run and among the smallest values measured in similar experiments. The image of Figure \ref{fig::divergence} (right)  shows instead the same image integrated over 30 laser shots which gives an overall cone of emission of approximately 10 mrad HWHM. The latter measurements gives an indication of the shot-to-shot pointing stability of the electron bunch. As for the origin of this fluctuation, it is unlikely to be affected by the laser pointing stability which was measured to be within the $\mu$rad range. The oscillations of the electron bunch inside the accelerating structure \cite{PhysRevLett.97.225002} is instead being explored as a possible explanation of the observed limit to the bunch pointing stability.

\begin{figure}[h]
\centering
\includegraphics[width=0.70\columnwidth]{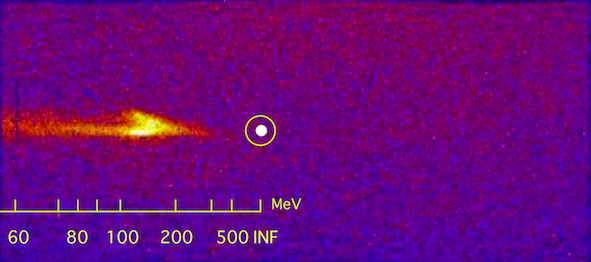}
\caption{Raw electron spectrum of a typical bunch accelerated from Nitrogen gas-jet. The spectrum shows a main component above 100 MeV and a low energy tail extending down to 60 MeV. The white dot indicates the landing position of undeflected electron.}
\label{fig::spectrum}
\end{figure}

Finally, information about the energy of the electron bunch was obtained by inserting the permanent magnet dipole downstream from the laser focal position. The image of Figure \ref{fig::spectrum} shows a typical spectrum of acceleration in Nitrogen in the same conditions of Figure \ref{fig::thomson}  and Figure \ref{fig::divergence} above. 
The white dot in the centre of the image indicates the average landing position of the electrons without the magnetic dipole. In the presence of the magnet, electrons will be deflected on the l.h.s. of the screen according to the dispersion curve of Figure \ref{fig::magnet_dispersion}. According to this image, the spectrum shows a main component around 150 MeV and a low energy tail extending down to 60 MeV. This general behavior was quite reproducible from shot to shot and accurate deconvolution of all the spectra is currently in progress.

\section{${\gamma}${}-RESIST conceptual set up\label{sec::gresist_setup}}
As shown in Figure \ref{fig::counterpropagating}, the experimental set up is based upon two counter-propagating laser pulses focused on a gas target using two off-axis parabolic mirrors. The two counter-propagating pulses are obtained by splitting the main laser pulse using a very thin splitter to minimize detrimental effects on the longitudinal phase of the transmitted pulse. The reflected pulse will be sent to the {\textquotedblleft}self-injection{\textquotedblright} arm (right) of the counter-propagating configuration, while the transmitted pulse will be sent to the {\textquotedblleft}Thomson/Compton scattering{\textquotedblright} arm. Given the small thickness of the splitter, it will be likely affected by flatness distortions which may induce phase distortions in the reflected pulse. The planned full beam adaptive optics will be used to compensate for this distortions and optimize both focal spots.
The \ {\textquotedblleft}self-injection{\textquotedblright} arm will be based upon the current set up and will use the existing 1 m focal length, F/10 off-axis parabolic mirror which will enable a maximum intensity on target up to 2$\times 10^{19}$ W/cm$^2$. The specific self-injection acceleration regime to be ultimately used in ${\gamma}${}-RESIST will depend upon the detailed full modeling currently in progress. However, the starting configuration will likely consist in the experimental configuration already tested during the self-injection test experiment (SITE) and summarized above.
The {\textquotedblleft}Thomson/Compton scattering{\textquotedblright} arm will consist of an off-axis parabolic mirror with a 0.5 m focal length,  F/5 numerical aperture which will enable a maximum intensity on target up to 4$\times 10^{19}$ W/cm$^2$. Optimization in the design of this arm will include the possibility to define the exact location in which overlapping of the laser pulse and electron bunch will occur.  A delay line on one of the two beam lines will be used to control the effective position of the scattering laser pulse relative to the accelerating pulse. This control will be crucial for the identification of the Thomson emission and will be used to explore coupling of the laser pulse with the electron bunch along its propagation trajectory from the injection point to the exit of the plasma. 

\subsection{Control of self-injection \label{sec::bunch_control}}
As described, in the typical experimental conditions of laser-plasma acceleration, its possible to evaluate the maximum energy gain in term of available power (in focus) and plasma density. In a simplified description we can say that energy gain increases as plasma density decreases as a result of longer accelerating distance \cite{Esarey:2009aa}. A preliminary test of this behavior was carried out during both runs described above.  The pressure scan carried out during the first run, at the maximum laser energy of 1 J, showed injection and acceleration for plasma density down to minimum gas-jet pressure of 7 bar in the case of Nitrogen. Similar measurements taken during the second run, at the higher laser energy of 2$\times 10^{19}$ W/cm$^2$ showed injection at pressures as low as 1 bar, corresponding to approximately $6\times 10^{17}$.  Ionization induced injection has been proposed as a possible solution to further enhance injection \cite{PhysRevLett.105.105003} and will be explored in future tests. 
Additional options to enhance the control of the injection process it to rely on the colliding laser pulses scheme \cite{PhysRevE.70.016402, Faure:2006aa} which exploits the large ponderomotive force associated to the beat-wave produced at the overlapping region of two counter-propagating pulses in order to pre-accelerate and inject a bunch of electrons into the bubble.  However, in this case, only a limited amount of charge is injected, typically around 10 pC, which makes this approach not suitable in view of an efficient radiation source. 

\section{Conclusions}
The progress of laser-plasma acceleration with self-injection is motivating the development of secondary radiation sources with unique properties, easily accessible for a wide range of applications. The $\gamma$-Resist project aims at demonstrating the generation of X/$\gamma$-rays using laser-accelerated electrons and Thomson/Compton scattering. Here an overview of the proposed scheme has been given, with attention to the expected performances and with a look at the possible use of the proposed experimental scheme for advanced studies of dynamics of electrons in intense fields. Finally, a preliminary presentation of recent experimental results on self-injection at FLAME was also given showing the achievement of effective acceleration of highly collimated, high energy electrons with moderate reproducibility and good pointing stability, a first step for the proposed radiation source.

\section{Acknowledgements}
 We thank the staff of the LNF Accelerator and Technical Divisions for the support during the the SITE operations at LNF.  "SITE" and "$\gamma$-RESIST" are funded by INFN through the CN5. The work was carried out in collaboration with the  High Field Photonics Unit (MD.P03.034) and X-ray Photonics (MD.P03.006.006) at INO-CNR  partially funded by CNR through the ELI-Italy project and by the MIUR-FIRB SPARX respectively. We acknowledge the CINECA Grant N. HP10CZX6QK2012 for the availability of high performance computing resources and the INFN APE project for the availability of the QUonG cluster.
\newline


\end{document}